\documentclass[twocolumn,showpacs,superscriptaddress,amssymb,10pt,pra,floatfix,aps,longbibliography]{revtex4-1}

\usepackage{graphicx}
\usepackage{dcolumn}
\usepackage{bm}
\usepackage{epsfig}
\usepackage{color}
\usepackage{longtable}
\usepackage{leftidx,amsmath}
\usepackage{natbib} 
\usepackage[utf8]{inputenc}

\def\ket#1{ $ \left\vert  #1   \right\rangle $ }

\def\mem#1#2#3{  \left\langle #1 \left\vert  #2 \right\vert #3 \right\rangle   }
\def\redme#1#2#3{ $ \left\langle #1 \left\Vert
	#2 \right\Vert #3 \right\rangle $ }
\def\redmem#1#2#3{  \left\langle #1 \left\Vert
	#2 \right\Vert #3 \right\rangle   }
\begin{document}
	\preprint{}
	%
%
%
%
	\title{Polarization effects in bound-free pair production}
%
%
%
%

	\author{J.~Sommerfeldt}
	\affiliation{Physikalisch--Technische Bundesanstalt, D--38116 Braunschweig, Germany}
	\affiliation{Technische Universit\"at Braunschweig, D--38106 Braunschweig, Germany}
	
	\author{R.~A.~M{\"u}ller}
	\affiliation{Physikalisch--Technische Bundesanstalt, D--38116 Braunschweig, Germany}
	\affiliation{Technische Universit\"at Braunschweig, D--38106 Braunschweig, Germany}

	\author{A.~N.~Artemyev}
	\affiliation{Institut f\"ur Physik und CINSaT, Universit\"at Kassel, D--34132 Kassel, Germany}
	
	\author{A.~Surzhykov}
	\affiliation{Physikalisch--Technische Bundesanstalt, D--38116 Braunschweig, Germany}
	\affiliation{Technische Universit\"at Braunschweig, D--38106 Braunschweig, Germany}	

	\date{\today \\[0.3cm]}

%
%
\begin{abstract}
We present a theoretical study of bound-free electron-positron pair production in the interaction of $\gamma$-rays with bare ions. Special attention is paid to the longitudinal polarization of both the emitted positrons and the produced hydrogen-like ions. To evaluate this polarization we employed exact solutions of the relativistic Dirac equation and treat the electron-photon coupling within the framework of first-order perturbation theory. Detailed calculations have been performed for both, low- and high-Z ions and for a wide range of photon energies. The results of these calculations suggest that bound-free pair production can be a source of strongly polarized positrons and ions.
\end{abstract}

\maketitle
	\section{Introduction}
		Owing to the development and construction of novel acceleration facilities such as FAIR in Darmstadt and Gamma-Factory at CERN \cite{krasny_cern_2018}, new interest arises to study high-energy ion-ion and ion-photon collisions. One of the most fundamental processes in these collisions is the creation of electron-positron pairs. The studies of this $e^-e^+$ process have a long history both in experiment and in theory. For example, due to its large cross section the creation of free $e^-e^+$ pairs has been extensively studied in ultra-relativistic ion-ion collisions \cite{overbo_exact_1968,budnev_two-photon_1975,ivanov_large_1999,lee_structure_2002,baur_electronpositron_2007}. Bound-free pair production is less probable in the high energy regime but still plays a significant role in accelerator physics since it leads to beam loss in heavy-ion colliders \cite{lab_conceptual_1989,bruce_observations_2007}. Theoretical analysis of this process can be performed very conveniently within the framework of the equivalent photon method by Weizs\"acker and Williams \cite{weizsacker_ausstrahlung_1934,williams_nature_1934}. In this approach the analysis of pair production in ion-ion collisions is traced back to its counterpart in photon-ion interactions. The investigation of photon induced pair production also attracts considerable attention since it allows us to gain more valuable information about light-matter interactions in the ultra-relativistic regime.

A large number of theoretical studies of photo-induced bound-free pair production has been performed during the last couple of decades \cite{aste_electromagnetic_1994,agger_pair_1997,belkacem_bound-free_1998,aste_bound-free_2008,deneke_bound-free_2008,artemyev_boundfree_2012}. Most of these studies have been focused on the total and angle-differential cross sections while much less attention has been paid to the polarization of the positrons and residual hydrogen-like ions. Detailed analysis of these polarization properties may help us to gain more insight into electron and positron dynamics in the relativistic regime. With the advance of positron spectrometers and storage ring techniques these studies become feasible, for example, in the FAIR and CERN facilities. In this work, therefore, we present a theoretical investigation of bound-free $e^-e^+$ pair production in collisions of $\gamma$-ray photons with bare ions. A special emphasis in this study is placed on the polarization of the produced positrons and hydrogen-like ions. In order to analyze these polarization properties we employ the relativistic Dirac equation to describe electron and positron states and first order perturbation theory for the coupling to the electromagnetic field. Based on this approach, in Sec. \ref{Theory} we derive the transition matrix element which is later used to calculate partial differential cross sections. By making use of these cross sections, we obtain the degree of polarization of the positrons and ions. Evaluation of these degrees of polarization requires high-demanding computations of free-bound integrals involving the radial components of the Dirac states. These computations are discussed in Sec. \ref{compDet}. Later in Sec. \ref{results}, we show the results of our calculations for interactions of photons in a wide range of energies with bare ions. In particular, we have found that bound-free pair production by circularly polarized light may lead to the production of strongly polarized positrons and residual hydrogen-like ions. Our results are finally summarized in Sec. \ref{Summary}. Relativistic units $\hbar = c = m_e = 1$ are used in this paper if not stated otherwise.
	\section{Theory} \label{Theory}
		\subsection{Evaluation of the transition amplitude} \label{TansMat}
			In relativistic theory, $e^-e^+$ pair production can be described as the excitation of an electron from the Dirac negative-energy continuum. The remaining hole in the Dirac sea corresponds to the produced positron. If during such an excitation the electron is captured into a bound ionic state, one talks about bound-free pair production. Analysis of all properties of this bound-free process can be traced back to the evaluation of the transition matrix element

\begin{equation} \label{IIMatEl}
M_{m_s\mu_f}(\lambda) = \int \text{d}\boldsymbol{r}~\psi{}^\dagger_{n_f\kappa_f\mu_f}(\boldsymbol{r}) \boldsymbol{\alpha} \cdot \boldsymbol{\hat{u}}_\lambda e^{i\boldsymbol{k}\boldsymbol{r}}\psi^{(+)}_{\boldsymbol{p},m_s}(\boldsymbol{r})~,
\end{equation}

\noindent where the coupling to the electromagnetic field is treated in Coulomb gauge and within the framework of first order perturbation theory. Evaluation of the matrix element~(\ref{IIMatEl}) requires explicit representations of the initial- and final-state wave functions, $\psi^{(+)}_{\boldsymbol{p},m_s}(\boldsymbol{r})$ and $\psi{}_{n_f\kappa_f\mu_f}(\boldsymbol{r})$, as well as the electron-photon interaction operator ${\hat{R} = \boldsymbol{\alpha} \cdot \boldsymbol{\hat{u}}_\lambda e^{i\boldsymbol{k}\boldsymbol{r}} }$. The wave function for the final state is given by the usual bound-electron solution of the Dirac equation 

\begin{equation} \label{IIBound}
\psi{}_{n_f\kappa{}_f\mu_f}(\boldsymbol{r}) = \left(\begin{array}{c}
g_{n_f\kappa{}_f}(r)\chi_{\kappa_f}^{\mu_f}(\boldsymbol{\hat{r}})\\
if_{n_f\kappa{}_f}(r)\chi_{-\kappa_f}^{\mu_f}(\boldsymbol{\hat{r}})\\
\end{array}\right)~,
\end{equation}

\noindent where $n_f$ is the principal quantum number, $\kappa_f$ is the Dirac quantum number and $\mu_f$ is the projection of the total angular momentum $j_f = \vert \kappa_f \vert - \frac{1}{2}$ \cite{eichler_relativistic_1995,eichler_lectures_2005}. In our work, this projection is defined with respect to the propagation direction of the incident light, which is chosen as the $z$-axis. Moreover in Eq. (\ref{IIBound}), $g_{n_f\kappa_f}(r)$ and $f_{n_f\kappa_f}(r)$ are the large and small radial components and $\chi_{\kappa_f}^{\mu_f}$ denotes the normalized spin-angular function.

In contrast to the bound-state wave function (\ref{IIBound}), $\psi^{(+)}_{\boldsymbol{p},m_s}(\boldsymbol{r})$ describes an electron in the Dirac negative-energy continuum. In scattering theory it is usually convenient to express this continuum solution as a decomposition into its partial waves. The explicit form of this multipole expansion depends on the choice of the axis along which the spin of the negative continuum electron is quantized. For proper analysis of polarization effects in $e^-e^+$ pair production, this axis has to be taken along the asymptotic momentum $\boldsymbol{p}$. In this so-called helicity representation, the electron wave function reads as

\begin{equation} \label{IICont}
\begin{aligned}
\psi^{(\pm)}_{\boldsymbol{p},m_s}(\boldsymbol{r}) = \sum_{\kappa{}_i\mu_i} i^{l_i} e^{\pm{}i\Delta_{\kappa_i}}\sqrt{4\pi(2l_i+1)} \langle l_i 0 \frac{1}{2} m_s\vert j_i m_s\rangle\\
\times \left(\begin{array}{c}
g_{E\kappa_i}(r)\chi_{\kappa_i}^{\mu_i}(\boldsymbol{\hat{r}})\\
if_{E\kappa_i}(r)\chi_{-\kappa_i}^{\mu_i}(\boldsymbol{\hat{r}})\\
\end{array}\right) D^{j_i}_{\mu_im_s}(\phi,\theta,0)~.
\end{aligned}
\end{equation}

\noindent Here, $m_s$ denotes the electron spin projection onto the propagation direction, $D^{j_i}_{\mu_im_s}(\phi,\theta,0)$ is the Wigner D-function, where $\phi$ and $\theta$ denote the azimuthal and polar angle of the electron asymptotic momentum, and

\begin{equation}
\Delta_{\kappa_i} = \delta_{\kappa_i} - \text{arg}\Gamma{}(s + i\eta) - \frac{1}{2}\pi{}s + (l_i+1)\frac{\pi}{2}
\end{equation}

\noindent is the difference between the asymptotic phases of the Dirac-Coulomb and free Dirac solutions. 

Eq. (\ref{IICont}) describes an electron with asymptotic momentum $\boldsymbol{p}$ in the Dirac negative-energy continuum. As already mentioned above, this wave function can be naturally used to describe the emitted positron. Namely, within the picture of the Dirac sea, the creation of an outgoing positron with energy $E_+ > 0$, momentum $\boldsymbol{p}_+$ and helicity $m_+$ is equivalent to the excitation of an incoming electron with energy $E = -E_+$, momentum $\boldsymbol{p} = -\boldsymbol{p}_+$ and helicity $m_s = m_+$. The radial components of such an negative-energy electron in a Coulomb potential are given by
\nopagebreak
\begin{equation} \label{IIradcomp}
\begin{aligned}
g_{E\kappa_i} (r) &= N_{\kappa_i} (\vert E \vert - 1)^\frac{1}{2}(2pr)^{s-1} \text{Re}[e^{-ipr}e^{i\delta_{\kappa_i}}(s + i\eta) \\
&\times \leftidx{_1}{F}{_1}(s+1+i\eta,2s+1;2ipr)]~,\\
f_{E\kappa_i} (r)&= N_{\kappa_i} (\vert E \vert + 1)^\frac{1}{2}(2pr)^{s-1} \text{Im}[e^{-ipr}e^{i\delta_{\kappa_i}}(s + i\eta) \\
&\times \leftidx{_1}{F}{_1}(s+1+i\eta,2s+1;2ipr)]~,\\
\end{aligned}
\end{equation}

\noindent with the parameters

\begin{equation}
\begin{aligned}
p = \sqrt{E^2 - 1},~\eta = \frac{\zeta E}{p},~\delta_{\kappa_i} = \frac{1}{2} \text{arg}\Big(\frac{-\kappa_i + i\eta/E}{s + i\eta}\Big),\\
\zeta = \alpha Z,~s = \sqrt{\kappa_i^2 - \zeta^2},~N_{\kappa_i} = 2\sqrt{\frac{p}{\pi}} e^{\pi\eta/2} \frac{\mid \Gamma{}(s + i\eta)\mid}{\Gamma{}(2s + 1)} .
\end{aligned}
\end{equation}

\noindent For more details see \cite{eichler_relativistic_1995,eichler_lectures_2005}.

So far, we have considered the initial- and final-state electron wave functions. The evaluation of the transition matrix element (\ref{IIMatEl}) also requires knowledge about the electron-photon interaction operator $\hat{R} = \boldsymbol{\alpha} \cdot \boldsymbol{\hat{u}}_\lambda e^{i\boldsymbol{k}\boldsymbol{r}}$. It is convenient to expand this operator in terms of the multipole components of the electromagnetic field \cite{rose_elementary_1957}. For light propagating in the $z$-direction, this expansion reads as

\begin{equation} \label{IIVector}
\hat{R} = \boldsymbol{\alpha}\cdot\boldsymbol{\hat{u}}_\lambda e^{ikz} = \sqrt{2\pi}\sum_{L=1}^\infty \sum_{p=0}^1 i^L\sqrt{2L+1}[(i\lambda)^p \boldsymbol{\alpha}\cdot\boldsymbol{a}^{(p)}_{L\lambda}]~,
\end{equation}

\noindent where $\boldsymbol{a}^{(0)}_{L\lambda}$ and $\boldsymbol{a}^{(1)}_{L\lambda}$ are the magnetic and electric multipole fields with angular momentum $L$. Moreover, $\lambda = \pm 1$ is the photon helicity.

Having discussed all the components of Eq. (\ref{IIMatEl}), we can further evaluate the transition matrix element $M_{m_s\mu_f}(\lambda)$. By inserting the wave functions (\ref{IIBound}) and (\ref{IICont}) into Eq. (\ref{IIMatEl}) we find

\begin{equation} \label{IIMatElHalf}
\begin{aligned}
M_{m_s\mu_f}(\lambda) &= \sum_{\kappa_i\mu_i} i^{l_i} e^{i\Delta_{\kappa_i}} \sqrt{4\pi(2l_i+1)} \langle l_i 0 \frac{1}{2} m_s\vert j_i m_s\rangle\\
\times &D^{j_i}_{\mu_im_s}(\phi,\theta,0) \mem{n_f\kappa_f\mu_f}{\boldsymbol{\alpha} \cdot \boldsymbol{\hat{u}}_\lambda e^{ikz}}{E\kappa_im_s},
\end{aligned}
\end{equation}

\noindent where 

\begin{equation} \label{IIMatElInt}
\begin{aligned}
&\mem{n_f\kappa_f\mu_f}{\boldsymbol{\alpha} \cdot \boldsymbol{\hat{u}}_\lambda e^{ikz}}{E\kappa_im_s}\\
=&\frac{i}{\sqrt{2}}\Bigg[\int \text{d}^3\boldsymbol{r} e^{i\boldsymbol{k}\boldsymbol{r}}g_{n_f\kappa{}_f}(r)(\chi_{\kappa_f}^{\mu_f})^\dagger (\sigma_x + i\lambda\sigma_y) f_{E\kappa_i}(r)\chi_{-\kappa_i}^{\mu_i}\\
&-\int \text{d}^3\boldsymbol{r} e^{i\boldsymbol{k}\boldsymbol{r}}f_{n_f\kappa{}_f}(r)(\chi_{-\kappa_f}^{\mu_f})^\dagger (\sigma_x + i\lambda\sigma_y) g_{E\kappa_i}(r)\chi_{\kappa_i}^{\mu_i}\Bigg]~.
\end{aligned}
\end{equation}

\noindent With the aid of the Wigner-Eckart theorem along with the multipole expansion (\ref{IIVector}) and after performing some simple algebra we finally obtain
\nopagebreak
\begin{equation} \label{IIMatElFull}
\begin{aligned}
M_{m_s\mu_f}(\lambda) &= \sqrt{8\pi^2}\sum_{\kappa_i\mu_iLp} i^{l_i+L} e^{i\Delta_{\kappa_i}} \sqrt{\frac{2l_i+1}{2j_f+1}}\sqrt{2L+1}\\
&\times \langle l_i 0 \frac{1}{2} m_s\vert j_i m_s\rangle \langle j_i \mu_i L \lambda\vert j_f \mu_f\rangle D^{j_i}_{\mu_im_s}(\phi,\theta,0)\\
&\times (i\lambda)^p\redmem{n_f\kappa_f}{\boldsymbol{\alpha}\cdot\boldsymbol{a}^{(p)}_L}{E\kappa_i}.
\end{aligned}
\end{equation}

\noindent Here, \redme{n_f\kappa_f}{\boldsymbol{\alpha}\cdot\boldsymbol{a}^{(p)}_L}{E\kappa_i} is the so-called reduced matrix element which is independent of the underlying geometry. It contains information about the electronic wave functions and is the central building block from which we calculate all properties of the process.
		\subsection{Differential cross sections and polarization parameters} \label{crossSec}
			In the previous section we have discussed the evaluation of the transition matrix element (\ref{IIMatEl}). With the help of this matrix element we can now analyse the angular and polarization properties of bound-free pair production. In particular, the angle-differential cross section of the process is obtained as

\begin{equation} \label{IICrossSec1}
\frac{\text{d}\sigma_{m_s\mu_f}}{\text{d}\Omega}(\lambda) = \frac{\alpha}{4k} \vert M_{m_s\mu_f}(\lambda) \vert ^2~,
\end{equation}

\noindent where we have assumed that the incident light has a well defined helicity $\lambda$ and the angular momentum projections of the outgoing positron and bound electron are observed. From cross section (\ref{IICrossSec1}), one can evaluate observables for all possible scenarios of pair production. For example, if the positron spin state remains unobserved in an experiment, the cross section is obtained by performing the summation over $m_s$, $\text{d}\sigma_{\mu_f} = \sum_{m_s} \text{d}\sigma_{m_s\mu_f}$. If, in contrast, no information about the magnetic sublevel population of the residual hydrogen-like ions is available, the cross section reads as $\text{d}\sigma_{m_s} = \sum_{\mu_f} \text{d}\sigma_{m_s\mu_f}$. Finally, if neither electron nor positron angular momentum projections are detected, we obtain the cross section

\begin{equation} \label{IICrossSec2}
\frac{\text{d}\sigma}{\text{d}\Omega}(\lambda) = \sum_{m_s\mu_f} \frac{\text{d}\sigma_{m_s\mu_f}}{\text{d}\Omega}(\lambda)~.
\end{equation}

\noindent Again, for all scenarios above, we assumed that the incident light is circularly polarized and hence has a well defined helicity $\lambda$. 

By using the differential cross sections introduced in this section, one can also calculate the degree of longitudinal polarization of the outgoing positrons and final hydrogen-like ions. For example, in a realistic experimental scenario in which the magnetic sublevel population of the ions remains unobserved, the degree of polarization of the positrons is given by
\nopagebreak
\begin{equation} \label{IIPosPolFunc}
P_{pos}(\theta) = \frac{\text{d}\sigma_{m_s=\frac{1}{2}} - \text{d}\sigma_{m_s=-\frac{1}{2}}}{\text{d}\sigma_{m_s=\frac{1}{2}} + \text{d}\sigma_{m_s=-\frac{1}{2}}}~.
\end{equation}

\noindent \textit{Vice versa}, the polarization of the hydrogen-like ions along the $z$-axis reads as

\begin{equation} \label{IIElPolFunc}
P_{ion}(\theta) = \frac{\text{d}\sigma_{\mu_f=\frac{1}{2}} - \text{d}\sigma_{\mu_f=-\frac{1}{2}}}{\text{d}\sigma_{\mu_f=\frac{1}{2}} + \text{d}\sigma_{\mu_f=-\frac{1}{2}}}~,
\end{equation}

\noindent where we assumed a positron detector that is insensitive to the spin state.
	\section{Computational Details} \label{compDet}
		The evaluation of matrix elements for bound-free transitions has been discussed many times in the literature not only in the context of pair production but also for the photoelectric effect \cite{pratt_atomic_1960, alling_exact_1965}, radiative recombination \cite{ichihara_radiative_1996} or single photon annihilation \cite{sodickson_single-quantum_1961}. For this reason, we restrict our discussion of the computational details to just a brief overview. As already mentioned above, the main building block of our analysis is the reduced matrix element in Eq. (\ref{IIMatElFull}) which consists of an angular and radial part. The angular part is given by the reduced matrix element of the spherical harmonic function and can be calculated analytically using the standard Racah algebra \cite{grant_relativistic_2007}. The radial part contains the integrals of the radial components of the electron and positron wave functions along with the spherical Bessel function of order $L$. Its numerical calculation is usually a rather complicated task. However, for $e^-e^+$ pair production in collisions of photons with initially bare ions, the radial components are known analytically \cite{eichler_relativistic_1995}. In this case, the exact solution of the radial integrals can be given in terms of Gaussian hypergeometric functions \cite{overbo_exact_1968} which we calculate numerically using the arb C library \cite{johansson_arb:_2017}. 

As seen from Eq. (\ref{IIMatElFull}) the reduced matrix elements for various positron and photon multipoles, $\kappa_i$ and $L$, contribute to $M_{m_s\mu_f}(\lambda)$. Since we consider a high energy atomic process a sufficiently large number of these partial waves have to be taken into account to achieve convergence of the cross section. For example, for $E_+ = 10 \text{~r.u.}$ our results contain partial waves up to $\vert \kappa_i \vert = 140$.
	\section{Results and Discussion} \label{results}
		\subsection{Differential cross sections} \label{crossSecRes}
\begin{figure*}
    \centering
      \includegraphics[width=0.75\textwidth]{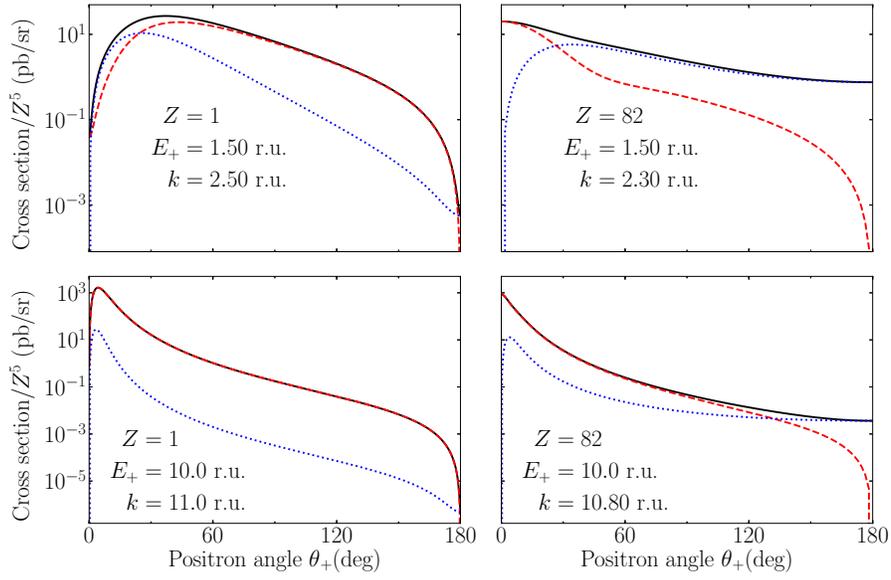}
       \caption{Differential cross sections for bound-free $e^-e^+$ pair production in collisions of photons with helicity $\lambda = +1$ and bare hydrogen (left panels) and lead (right panels) ions. Calculations have been performed for positron energies $E_+ = 1.5\text{~r.u.}$ (upper panels) and $E_+ = 10\text{~r.u.}$ (lower panels) and for capture of the electron in the ground ionic state. Moreover, three scenarios are considered in which the spin projection of the emitted positrons either remains unobserved (black solid line) or is fixed to $m_s = + 1/2$ (red dashed line) or $m_s = -1/2$ (blue dotted line). \label{IVPosCrossSection}}
\end{figure*}

			With the help of Eqs. (\ref{IIMatElFull}) - (\ref{IIElPolFunc}) we are now ready to investigate $e^-e^+$ bound-free pair production in photon-ion interactions. We start our analysis with the angle-differential cross sections which we calculate for photons with helicity $\lambda = +1$ colliding with bare hydrogen and lead ions. For both targets we focus on the capture of the produced electrons into the ground $1\text{s}_{1/2}$ ionic state and consider low and high energy regimes corresponding to positron energies of $E_+ = 1.5 \text{~r.u.}$ and $E_+ = 10 \text{~r.u.}$, respectively. Moreover, we discuss different "polarization scenarios" in which the angular momentum projections of either the final hydrogen-like ions or emitted positrons are observed. In Fig. \ref{IVPosCrossSection} for example, we display the differential cross section $\text{d}\sigma_{m_s} = \sum_{\mu_f} \text{d}\sigma_{m_s\mu_f}$ which is obtained upon summation over the final ionic states but in which the positron helicity is fixed to $m_s = + 1/2$ (red dashed line) or $m_s = - 1/2$ (blue dotted line). The sum of the two helicity contributions $\text{d}\sigma = \sum_{m_s}\text{d}\sigma_{m_s}$ is displayed by the black solid line. As seen from the figure, this summed angle-differential cross section $\text{d}\sigma$ behaves in a rather different way for light (hydrogen) and heavy (lead) target ions. In particular, while for $Z = 82$ the maximum of positron emission is in the forward direction, $\theta_+ = 0^\circ$, it is shifted to higher angles $\theta_+$ for $Z = 1$. For example for positrons with energy $E_+ = 1.5 \text{~r.u.}$, the differential cross section $\text{d}\sigma$ has its maximum at $\theta_+ \approx 37^\circ$. This behaviour has been previously predicted within the framework of the relativistic Born approximation in which the outgoing positron is treated as a plane wave \cite{pratt_atomic_1960,agger_pair_1997,bhabha_h._j._annihilation_1934,johnson_angular_1967}. The Born approximation also suggests the drastic suppression of positron emission in the forward and backward directions in the low $Z$ regime; this can be clearly observed in the left panels of Fig. \ref{IVPosCrossSection}.

The angular distribution of the emitted positrons also depends on their helicity $m_s$. It is particularly easy to see the effect for the lead target for which the forward emission is dominated by positrons with helicity $m_s = +1/2$ while positrons with $m_s = - 1/2$ are most likely emitted under large angles. This behaviour can be understood based on the analysis of the angular momentum projections for the two ultimate cases $\theta_+ = 0^\circ$ and $\theta_+ = 180^\circ$, \textit{i.e.} for propagation of the positron either parallel or antiparallel to the $z$-axis. For these two cases the spin projections of the bound-electron and emitted positron should add up to the helicity of the incident photon, 

\begin{equation} \label{IVAngCons}
\lambda = \mu_f + \widetilde{m_s}~.
\end{equation}

\noindent Here, $\widetilde{m_s}$ is \textit{not} the helicity but the projection of the positron spin on the $z$-axis. If the electron is captured into the ground ionic state its angular momentum projection can be $\mu_f = \pm 1/2$. Therefore, only the combination $\mu_f = +1/2$ and $\widetilde{m_s} = +1/2$ can compensate the helicity $\lambda = +1$, see Eq. (\ref{IVAngCons}). However as we already mentioned above, $\widetilde{m_s}$ is the projection of the positron spin on the direction of the incident light which is related to the helicity as $m_s = \widetilde{m_s} = +1/2$ for $\theta_+ = 0^\circ$ and $m_s = -\widetilde{m_s} = -1/2$ for $\theta_+ = 180^\circ$.

\begin{figure*}
    \centering
      \includegraphics[width=0.75\textwidth]{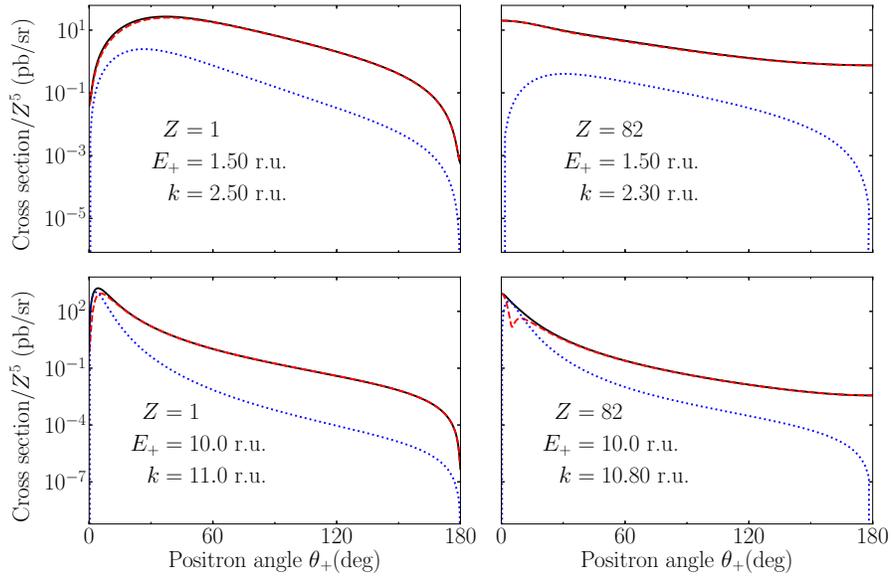}
       \caption{As in Fig. \ref{IVPosCrossSection} but for well defined magnetic quantum numbers $\mu_f = +1/2$ (red dashed line) and $\mu_f = -1/2$ (blue dotted line) of the bound-state electron while the positron spin state remains unobserved. As before, the black solid line represents the summed partial differential cross section.\label{IVElCrossSection}}
\end{figure*}

Until now we have discussed the differential cross section for bound-free pair production under the assumption that the spin state of the final hydrogen-like ions remains unobserved. In order to investigate how the probability of the $e^-e^+$ process depends on the magnetic sublevel population of the residual ions, we display in Fig. \ref{IVElCrossSection} the differential cross section $\text{d}\sigma_{\mu_f} = \sum_{m_s} \text{d}\sigma_{m_s\mu_f}$ for $\mu_f = +1/2$ (red dashed line) and $\mu_f = -1/2$ (blue dotted line). This cross section has been obtained upon summation over the positron spin states and for incident photons with helicity $\lambda = +1$. Similar to before, calculations have been performed for hydrogen and lead ions as well as for positron energies $E_+ = 1.5 \text{~r.u.}$  and $E_+ = 10 \text{~r.u.}$. As seen from the figure, for both energies and targets, and nearly for all emission angles, the $e^-e^+$ process leads almost exclusively to the production of hydrogen-like ions with the magnetic quantum number $\mu_f = + 1/2$. To explain this effect one has to revisit Eq. (\ref{IIMatElInt}) in which the matrix element $\mem{n_f\kappa_f\mu_f}{\boldsymbol{\alpha} \cdot \boldsymbol{\hat{u}}_\lambda e^{ikz}}{E\kappa_im_s}$ is written as the sum of two integrals. The first integral contains the product of the upper electron $g_{n_f\kappa_f}$ and lower positron $f_{E\kappa_i}$ components. In contrast, the product of the lower electron $f_{n_f\kappa_f}$ and upper positron $g_{E\kappa_i}$ functions can be found under the second integral. For moderate relativistic energies the contribution of this second integral to the matrix element is negligible since both $f_{n_f\kappa_f}$ and $g_{E\kappa_i}$ are \textit{small} components. Therefore, the behaviour of the differential cross is mainly determined by the first term in Eq. (\ref{IIMatElInt}) which includes the \textit{large} electron and positron components. However, the angular part of this integral $(\chi_{\kappa_f}^{\mu_f})^\dagger (\sigma_x + i\lambda\sigma_y)\chi_{-\kappa_i}^{\mu_i}$ vanishes for $\mu_f = -1/2$ and $\lambda = +1$. This matches our observation from Fig. \ref{IVElCrossSection} of the strongly suppressed production of ions with $\mu_f = -1/2$. As seen from Eq. (\ref{IIradcomp}), for very high energies $\vert E \vert >> 1$ the upper and lower components of the positron wave function become comparable which leads to the enhancement of the contribution of the second integral in Eq. (\ref{IIMatElInt}). This integral doesn't disappear for $\lambda = +1$ and $\mu_f = -1/2$ and can result in a significant contribution of $\text{d}\sigma_{\mu_f = -1/2}$. As seen from the lower panels of Fig. \ref{IVElCrossSection}, this partial cross section even becomes dominant for a small range of emission angles around $\theta_+ \approx 2.5^\circ$ for $Z = 1$ and $\theta_+ \approx 6.5^\circ$ for $Z = 82$.

		\subsection{Degree of polarization} \label{polDeg}
\begin{figure}
    \centering
      \includegraphics[width=0.9\linewidth]{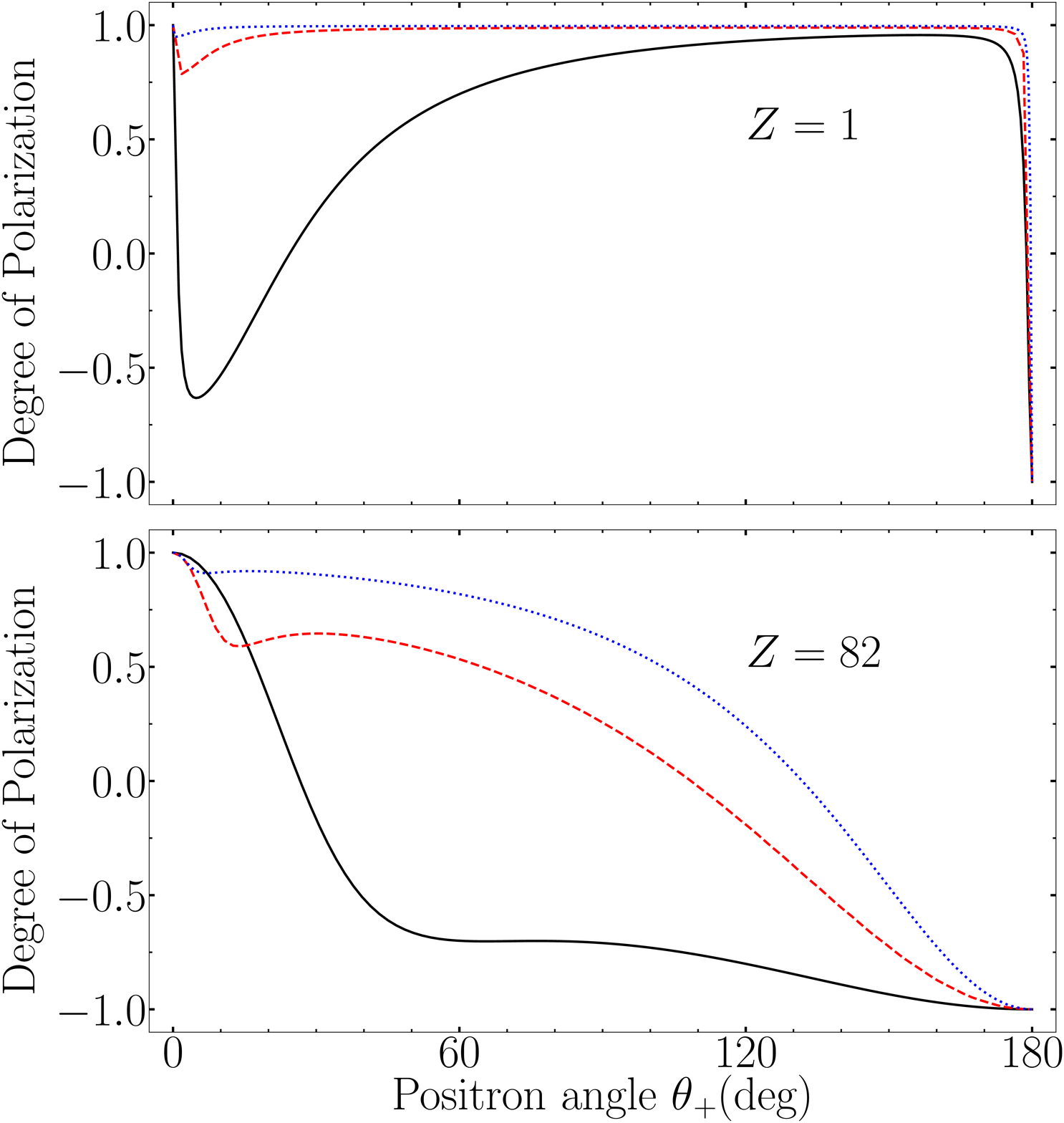}
      \caption{Degree of polarization of the created positrons (\ref{IIPosPolFunc}) for bound-free $e^-e^+$ pair production in collisions of photons with helicity $\lambda = +1$ and bare hydrogen (upper panel) and lead (lower panel) ions.  Calculations have been performed for positron energies $E_+ = 1.5\text{~r.u.}$ (black solid line), $E_+ = 5\text{~r.u.}$ (red dashed line) and $E_+ = 10\text{~r.u.}$ (blue dotted line) and for capture of the electron into the ground $1\text{s}_{1/2}$ ionic state.\label{IVPosPol}}
\end{figure}

			As we have discussed in the previous section, the process of bound-free pair production is very sensitive to the spin state of the emitted positrons and residual hydrogen-like ions. In order to investigate this $m_s$- and $\mu_f$-dependence in detail it is convenient to analyse not only the partial differential cross sections but also the degrees of positron and ion polarization, Eqs. (\ref{IIPosPolFunc}) and (\ref{IIElPolFunc}). Similar to before we start with the emitted positrons whose degree of polarization $P_{pos}(\theta_+)$ is displayed in Fig. \ref{IVPosPol}. To be consistent with the results of the previous section, calculations have been performed for collisions of photons with helicity $\lambda = +1$ with bare hydrogen and lead ions as well as for positron energies $E_+ = 1.5\text{~r.u.}$, $E_+ = 5\text{~r.u.}$ and $E_+ = 10\text{~r.u.}$. Moreover, $P_{pos}(\theta_+)$ has been obtained under the assumption that the spin state of the hydrogen-like ions remains unobserved. As seen from the figure, $P_{pos}(\theta_+) \approx +1$ for $\theta_+ \to 0^\circ$ implying that for the forward emission positrons are strongly polarized in the direction of propagation. In contrast, for larger angles $\theta_+$ the degree of polarization decreases and reaches the value $P_{pos}(\theta_+) = -1$ for $\theta_+ = 180^\circ$; the effect which can be expected from Eq. (\ref{IVAngCons}). The behaviour of $P_{pos}(\theta_+)$ between the two ultimate angles $\theta_+ = 0^\circ$ and $\theta_+ = 180^\circ$ strongly depends on the positron energy and charge of the target ion. For very high relativistic energies for example, the creation of positrons with helicity $m_s = + 1/2$ remains dominant in a rather large angular range. This effect is most pronounced for hydrogen ions and positron energies of $E_+ = 10\text{~r.u.}$ for which $P_{pos}(\theta_+) \approx +1$ for $0^\circ \leq  \theta_+ \lesssim 178^\circ$.

One may note from Fig. \ref{IVPosPol} that our results for the degree of positron polarization differ from those of Agger and S\o{}rensen \cite{agger_pair_1997}. The reason for this disagreement is the choice of the axis with respect to which the spin of the created positrons is quantized. In reference \cite{agger_pair_1997} this axis was chosen along the propagation direction of the incident light. However, this choice is insufficient to analyse the polarization effects in pair production because for relativistic particles the only direction along which one can uniquely define polarization is their own direction of propagation. Therefore, one has to define the partial differential cross sections in Eq. (\ref{IIPosPolFunc}) in the helicity basis, \textit{i.e.} with respect to the asymptotic momentum of the created positrons $\boldsymbol{p}_+$. Such calculations are shown in Fig.~\ref{IVPosPol}.

Besides the emitted positrons it is also instructive to analyse the degree of polarization of the produced hydrogen-like ions $P_{ion}(\theta_+)$. It is obtained from Eq. (\ref{IIElPolFunc}) where the magnetic quantum number $\mu_f$ is defined with respect to the propagation direction of the incident light. Predictions for $P_{ion}(\theta_+)$ are presented in Fig. 4 for the same set of parameters as used above for the positron polarization. The figure clearly indicates that for positron emission in the forward and backward direction the hydrogen-like ions are always produced in the magnetic substate \ket{1\text{s}_{1/2},\mu_f = +1/2}. Moreover, almost exclusive population of the state with $\mu_f = +1/2$ can be observed for relatively low positron energies (see black solid line). Only in the strongly relativistic regime, the predominant population of the sublevel with $\mu_f = -1/2$ becomes possible in a rather restricted range of forward emission angles. For example for $E_+ = 10\text{~r.u.}$, the partial cross section $\text{d}\sigma_{\mu_f = -1/2}$ is dominant for the range $0.2^\circ \lesssim \theta_+ \lesssim 4.7^\circ$ for $Z = 1$ and $2.5^\circ \lesssim \theta_+ \lesssim 10.7^\circ$ for $Z = 82$. A detailed explanation of this effect based on the analysis of Eq. (\ref{IIMatElInt}) has been given in the previous section. The results of our calculations, therefore, indicate that bound-free pair production by circularly polarized light may be used to produce hydrogen-like ions with a high degree of longitudinal polarization.

\begin{figure}
    \centering
      \includegraphics[width=0.9\linewidth]{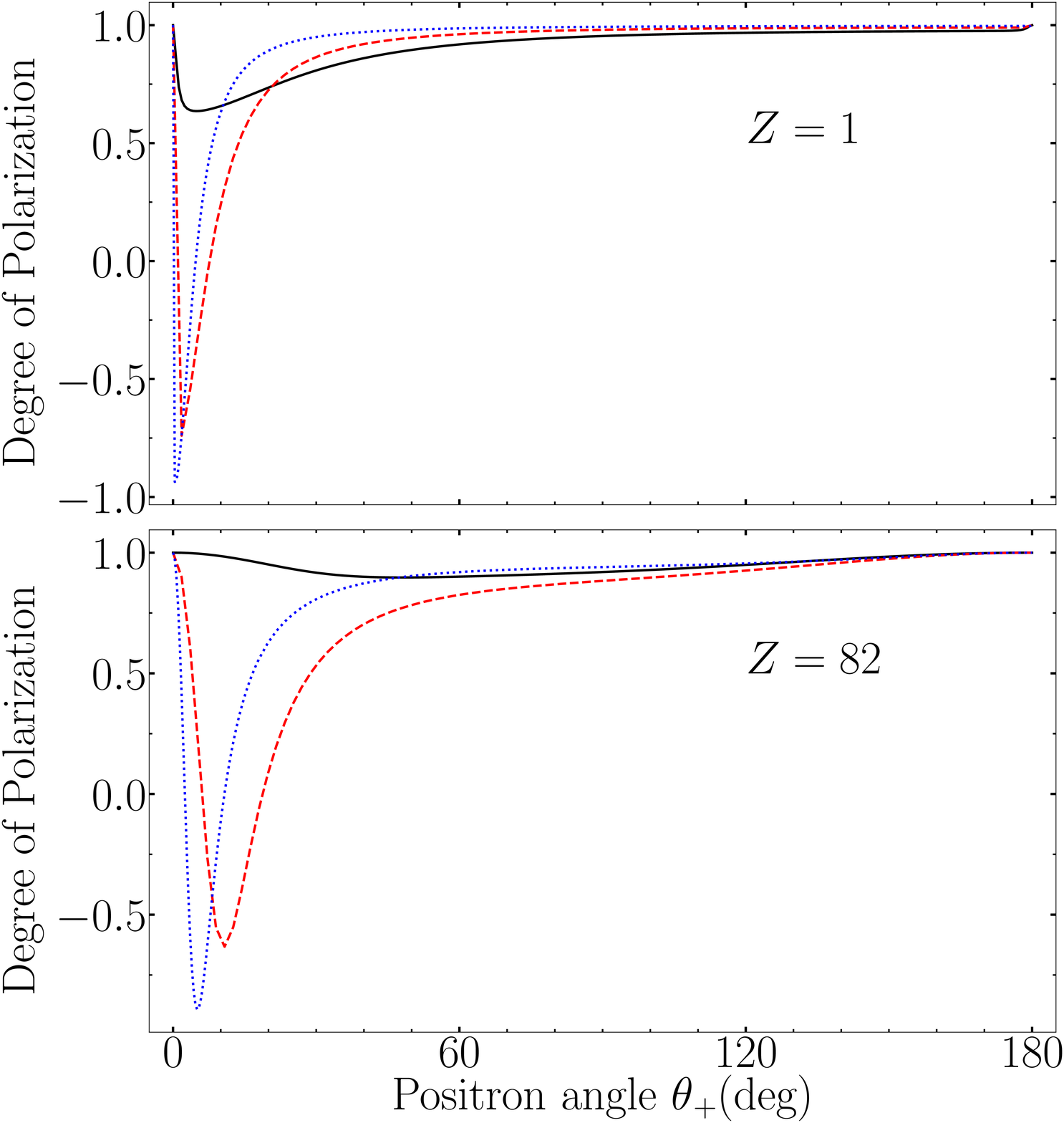}
      \caption{Degree of polarization of the residual hydrogen-like ions (\ref{IIElPolFunc}) for the same parameters as in Fig. \ref{IVPosPol}.\label{IVElPol}}
\end{figure}
	\section{Summary} \label{Summary}
			In conclusion, we presented a theoretical study of bound-free electron-positron pair production in the interaction of $\gamma$-rays with bare ions. Based on the rigorous solutions of the relativistic Dirac equation and by making use of the first-order perturbation theory for the electron-photon coupling, we studied the polarization properties of the created positrons and residual hydrogen-like ions. Calculations have been performed for circularly polarized photons in a wide range of energies and for low- and high-$Z$ targets. Results of these calculations have clearly shown that in the relatively low energy regime, \textit{i.e.} when $E_+ \approx m_e c^2$, the produced ions are strongly longitudinally polarized in the direction of the incident photon beam for all positron emission angles. In contrast, for high energies $E_+ >> m_e c^2$ and forward positron emission $\theta_+ > 0$, the ions are polarized opposite to the photon wave vector. Moreover, also the positron spin state strongly depends on the energy of the incident light and nuclear charge of the target. For example for low-$Z$ targets and ultra relativistic energies, the positrons are almost exclusively created in the spin state with $m_s = + 1/2$. We argue, therefore, that bound-free pair production can be used as a source of strongly polarized positrons and hydrogen-like ions; this effect is likely to be observed soon in the future FAIR facility in Darmstadt and in the Gamma-Factory at CERN.
	\bibliography{atom}

\begin{thebibliography}{27}%
\makeatletter
\providecommand \@ifxundefined [1]{%
 \@ifx{#1\undefined}
}%
\providecommand \@ifnum [1]{%
 \ifnum #1\expandafter \@firstoftwo
 \else \expandafter \@secondoftwo
 \fi
}%
\providecommand \@ifx [1]{%
 \ifx #1\expandafter \@firstoftwo
 \else \expandafter \@secondoftwo
 \fi
}%
\providecommand \natexlab [1]{#1}%
\providecommand \enquote  [1]{``#1''}%
\providecommand \bibnamefont  [1]{#1}%
\providecommand \bibfnamefont [1]{#1}%
\providecommand \citenamefont [1]{#1}%
\providecommand \href@noop [0]{\@secondoftwo}%
\providecommand \href [0]{\begingroup \@sanitize@url \@href}%
\providecommand \@href[1]{\@@startlink{#1}\@@href}%
\providecommand \@@href[1]{\endgroup#1\@@endlink}%
\providecommand \@sanitize@url [0]{\catcode `\\12\catcode `\$12\catcode
  `\&12\catcode `\#12\catcode `\^12\catcode `\_12\catcode `\%12\relax}%
\providecommand \@@startlink[1]{}%
\providecommand \@@endlink[0]{}%
\providecommand \url  [0]{\begingroup\@sanitize@url \@url }%
\providecommand \@url [1]{\endgroup\@href {#1}{\urlprefix }}%
\providecommand \urlprefix  [0]{URL }%
\providecommand \Eprint [0]{\href }%
\providecommand \doibase [0]{http://dx.doi.org/}%
\providecommand \selectlanguage [0]{\@gobble}%
\providecommand \bibinfo  [0]{\@secondoftwo}%
\providecommand \bibfield  [0]{\@secondoftwo}%
\providecommand \translation [1]{[#1]}%
\providecommand \BibitemOpen [0]{}%
\providecommand \bibitemStop [0]{}%
\providecommand \bibitemNoStop [0]{.\EOS\space}%
\providecommand \EOS [0]{\spacefactor3000\relax}%
\providecommand \BibitemShut  [1]{\csname bibitem#1\endcsname}%
\let\auto@bib@innerbib\@empty
\bibitem [{\citenamefont {Krasny}\ \emph {et~al.}(2018)\citenamefont {Krasny},
  \citenamefont {Alemany-Fern\'andez}, \citenamefont {Antsifarov},
  \citenamefont {Apyan}, \citenamefont {Bartosik}, \citenamefont {Bessonov},
  \citenamefont {Biancacci}, \citenamefont {Bieron}, \citenamefont {Budker},
  \citenamefont {Cassou}, \citenamefont {Castelli}, \citenamefont {Chaikovska},
  \citenamefont {Chehab}, \citenamefont {Curatolo}, \citenamefont {Czodrowski},
  \citenamefont {Dupraz}, \citenamefont {Dzierzega}, \citenamefont {Goddard},
  \citenamefont {Hirlaender}, \citenamefont {Jowett}, \citenamefont {Kersevan},
  \citenamefont {Kowalska}, \citenamefont {Kroeger}, \citenamefont {Lamont},
  \citenamefont {Manglunki}, \citenamefont {Martens}, \citenamefont {Petrenko},
  \citenamefont {Petrillo}, \citenamefont {Placzek}, \citenamefont {Pustelny},
  \citenamefont {Schaumann}, \citenamefont {Serafini}, \citenamefont
  {Shevelko}, \citenamefont {St\"ohlker}, \citenamefont {Weber}, \citenamefont
  {Wu}, \citenamefont {Yin~Vallgren}, \citenamefont {Zimmermann}, \citenamefont
  {Zolotorev},\ and\ \citenamefont {Zomer}}]{krasny_cern_2018}%
  \BibitemOpen
  \bibfield  {author} {\bibinfo {author} {\bibfnamefont {Mieczyslaw}\
  \bibnamefont {Krasny}}, \bibinfo {author} {\bibfnamefont {Reyes}\
  \bibnamefont {Alemany-Fern\'andez}}, \bibinfo {author} {\bibfnamefont
  {P.}~\bibnamefont {Antsifarov}}, \bibinfo {author} {\bibfnamefont {Armen}\
  \bibnamefont {Apyan}}, \bibinfo {author} {\bibfnamefont {Hannes}\
  \bibnamefont {Bartosik}}, \bibinfo {author} {\bibfnamefont {Evgeny}\
  \bibnamefont {Bessonov}}, \bibinfo {author} {\bibfnamefont {Nicolo}\
  \bibnamefont {Biancacci}}, \bibinfo {author} {\bibfnamefont {Jacek}\
  \bibnamefont {Bieron}}, \bibinfo {author} {\bibfnamefont {Dmitry}\
  \bibnamefont {Budker}}, \bibinfo {author} {\bibfnamefont {Kevin}\
  \bibnamefont {Cassou}}, \bibinfo {author} {\bibfnamefont {Fabrizio}\
  \bibnamefont {Castelli}}, \bibinfo {author} {\bibfnamefont {Iryna}\
  \bibnamefont {Chaikovska}}, \bibinfo {author} {\bibfnamefont {Robert}\
  \bibnamefont {Chehab}}, \bibinfo {author} {\bibfnamefont {Camilla}\
  \bibnamefont {Curatolo}}, \bibinfo {author} {\bibfnamefont {Patrick}\
  \bibnamefont {Czodrowski}}, \bibinfo {author} {\bibfnamefont {Kevin}\
  \bibnamefont {Dupraz}}, \bibinfo {author} {\bibfnamefont {Krzysztof}\
  \bibnamefont {Dzierzega}}, \bibinfo {author} {\bibfnamefont {Brennan}\
  \bibnamefont {Goddard}}, \bibinfo {author} {\bibfnamefont {Simon}\
  \bibnamefont {Hirlaender}}, \bibinfo {author} {\bibfnamefont {John}\
  \bibnamefont {Jowett}}, \bibinfo {author} {\bibfnamefont {Roberto}\
  \bibnamefont {Kersevan}}, \bibinfo {author} {\bibfnamefont {Magdalena}\
  \bibnamefont {Kowalska}}, \bibinfo {author} {\bibfnamefont {Felix}\
  \bibnamefont {Kroeger}}, \bibinfo {author} {\bibfnamefont {Mike}\
  \bibnamefont {Lamont}}, \bibinfo {author} {\bibfnamefont {Django}\
  \bibnamefont {Manglunki}}, \bibinfo {author} {\bibfnamefont {Aurélien}\
  \bibnamefont {Martens}}, \bibinfo {author} {\bibfnamefont {Alexey}\
  \bibnamefont {Petrenko}}, \bibinfo {author} {\bibfnamefont {Vittoria}\
  \bibnamefont {Petrillo}}, \bibinfo {author} {\bibfnamefont {Wieslaw}\
  \bibnamefont {Placzek}}, \bibinfo {author} {\bibfnamefont {Szymon}\
  \bibnamefont {Pustelny}}, \bibinfo {author} {\bibfnamefont {Michaela}\
  \bibnamefont {Schaumann}}, \bibinfo {author} {\bibfnamefont {Luca}\
  \bibnamefont {Serafini}}, \bibinfo {author} {\bibfnamefont {Viacheslaw}\
  \bibnamefont {Shevelko}}, \bibinfo {author} {\bibfnamefont {Thomas}\
  \bibnamefont {St\"ohlker}}, \bibinfo {author} {\bibfnamefont {Guenter}\
  \bibnamefont {Weber}}, \bibinfo {author} {\bibfnamefont {Ying}\ \bibnamefont
  {Wu}}, \bibinfo {author} {\bibfnamefont {Christina}\ \bibnamefont
  {Yin~Vallgren}}, \bibinfo {author} {\bibfnamefont {Frank}\ \bibnamefont
  {Zimmermann}}, \bibinfo {author} {\bibfnamefont {Max}\ \bibnamefont
  {Zolotorev}}, \ and\ \bibinfo {author} {\bibfnamefont {Fabian}\ \bibnamefont
  {Zomer}},\ }\bibfield  {title} {\enquote {\bibinfo {title} {The {CERN}
  {Gamma} {Factory} {Initiative}: {An} {Ultra}-{High} {Intensity} {Gamma}
  {Source}},}\ }\href@noop {} {\bibfield  {journal} {\bibinfo  {journal}
  {Proceedings of the 9th Int. Particle Accelerator Conf.}\ }\textbf {\bibinfo
  {volume} {IPAC2018}} (\bibinfo {year} {2018})}\BibitemShut {NoStop}%
\bibitem [{\citenamefont {{\O}verb{\o}}\ \emph {et~al.}(1968)\citenamefont
  {{\O}verb{\o}}, \citenamefont {Mork},\ and\ \citenamefont
  {A.~Olsen}}]{overbo_exact_1968}%
  \BibitemOpen
  \bibfield  {author} {\bibinfo {author} {\bibfnamefont {Ingjald}\ \bibnamefont
  {{\O}verb{\o}}}, \bibinfo {author} {\bibfnamefont {Kjell}\ \bibnamefont
  {Mork}}, \ and\ \bibinfo {author} {\bibfnamefont {Haakon}\ \bibnamefont
  {A.~Olsen}},\ }\bibfield  {title} {\enquote {\bibinfo {title} {Exact
  {Calculation} of {Pair} {Production}},}\ }\href {\doibase
  10.1103/PhysRev.175.1978} {\bibfield  {journal} {\bibinfo  {journal}
  {Physical Review}\ }\textbf {\bibinfo {volume} {175}},\ \bibinfo {pages}
  {1978--1981} (\bibinfo {year} {1968})}\BibitemShut {NoStop}%
\bibitem [{\citenamefont {Budnev}\ \emph {et~al.}(1975)\citenamefont {Budnev},
  \citenamefont {Ginzburg}, \citenamefont {Meledin},\ and\ \citenamefont
  {Serbo}}]{budnev_two-photon_1975}%
  \BibitemOpen
  \bibfield  {author} {\bibinfo {author} {\bibfnamefont {V.~M.}\ \bibnamefont
  {Budnev}}, \bibinfo {author} {\bibfnamefont {I.~F.}\ \bibnamefont
  {Ginzburg}}, \bibinfo {author} {\bibfnamefont {G.~V.}\ \bibnamefont
  {Meledin}}, \ and\ \bibinfo {author} {\bibfnamefont {V.~G.}\ \bibnamefont
  {Serbo}},\ }\bibfield  {title} {\enquote {\bibinfo {title} {The two-photon
  particle production mechanism. {Physical} problems. {Applications}.
  {Equivalent} photon approximation},}\ }\href {\doibase
  10.1016/0370-1573(75)90009-5} {\bibfield  {journal} {\bibinfo  {journal}
  {Physics Reports}\ }\textbf {\bibinfo {volume} {15}},\ \bibinfo {pages}
  {181--282} (\bibinfo {year} {1975})}\BibitemShut {NoStop}%
\bibitem [{\citenamefont {Ivanov}\ \emph {et~al.}(1999)\citenamefont {Ivanov},
  \citenamefont {Schiller},\ and\ \citenamefont {Serbo}}]{ivanov_large_1999}%
  \BibitemOpen
  \bibfield  {author} {\bibinfo {author} {\bibfnamefont {D.~Yu.}\ \bibnamefont
  {Ivanov}}, \bibinfo {author} {\bibfnamefont {A.}~\bibnamefont {Schiller}}, \
  and\ \bibinfo {author} {\bibfnamefont {V.~G.}\ \bibnamefont {Serbo}},\
  }\bibfield  {title} {\enquote {\bibinfo {title} {Large {Coulomb} corrections
  to the $e^+e^-$ pair production at relativistic heavy ion colliders},}\
  }\href {\doibase 10.1016/S0370-2693(99)00323-8} {\bibfield  {journal}
  {\bibinfo  {journal} {Physics Letters B}\ }\textbf {\bibinfo {volume}
  {454}},\ \bibinfo {pages} {155--160} (\bibinfo {year} {1999})}\BibitemShut
  {NoStop}%
\bibitem [{\citenamefont {Lee}\ \emph {et~al.}(2002)\citenamefont {Lee},
  \citenamefont {Milstein},\ and\ \citenamefont {Serbo}}]{lee_structure_2002}%
  \BibitemOpen
  \bibfield  {author} {\bibinfo {author} {\bibfnamefont {R.~N.}\ \bibnamefont
  {Lee}}, \bibinfo {author} {\bibfnamefont {A.~I.}\ \bibnamefont {Milstein}}, \
  and\ \bibinfo {author} {\bibfnamefont {V.~G.}\ \bibnamefont {Serbo}},\
  }\bibfield  {title} {\enquote {\bibinfo {title} {Structure of the {Coulomb}
  and unitarity corrections to the cross section of $e^+e^-$ pair production in
  ultrarelativistic nuclear collisions},}\ }\href {\doibase
  10.1103/PhysRevA.65.022102} {\bibfield  {journal} {\bibinfo  {journal}
  {Physical Review A}\ }\textbf {\bibinfo {volume} {65}},\ \bibinfo {pages}
  {022102} (\bibinfo {year} {2002})}\BibitemShut {NoStop}%
\bibitem [{\citenamefont {Baur}\ \emph {et~al.}(2007)\citenamefont {Baur},
  \citenamefont {Hencken},\ and\ \citenamefont
  {Trautmann}}]{baur_electronpositron_2007}%
  \BibitemOpen
  \bibfield  {author} {\bibinfo {author} {\bibfnamefont {Gerhard}\ \bibnamefont
  {Baur}}, \bibinfo {author} {\bibfnamefont {Kai}\ \bibnamefont {Hencken}}, \
  and\ \bibinfo {author} {\bibfnamefont {Dirk}\ \bibnamefont {Trautmann}},\
  }\bibfield  {title} {\enquote {\bibinfo {title} {Electron–positron pair
  production in ultrarelativistic heavy ion collisions},}\ }\href {\doibase
  10.1016/j.physrep.2007.09.002} {\bibfield  {journal} {\bibinfo  {journal}
  {Physics Reports}\ }\textbf {\bibinfo {volume} {453}},\ \bibinfo {pages}
  {1--27} (\bibinfo {year} {2007})}\BibitemShut {NoStop}%
\bibitem [{\citenamefont {Lab}(1989)}]{lab_conceptual_1989}%
  \BibitemOpen
  \bibfield  {author} {\bibinfo {author} {\bibfnamefont {Brookhaven~National}\
  \bibnamefont {Lab}},\ }\href@noop {} {\emph {\bibinfo {title} {Conceptual
  design of the {Relativistic} {Heavy} {Ion} {Collider} [{RHIC}]}}},\ \bibinfo
  {type} {Tech. Rep.}\ \bibinfo {number} {BNL–52195}\ (\bibinfo
  {institution} {Brookhaven National Lab.},\ \bibinfo {year}
  {1989})\BibitemShut {NoStop}%
\bibitem [{\citenamefont {Bruce}\ \emph {et~al.}(2007)\citenamefont {Bruce},
  \citenamefont {Jowett}, \citenamefont {Gilardoni}, \citenamefont {Drees},
  \citenamefont {Fischer}, \citenamefont {Tepikian},\ and\ \citenamefont
  {Klein}}]{bruce_observations_2007}%
  \BibitemOpen
  \bibfield  {author} {\bibinfo {author} {\bibfnamefont {R.}~\bibnamefont
  {Bruce}}, \bibinfo {author} {\bibfnamefont {J.~M.}\ \bibnamefont {Jowett}},
  \bibinfo {author} {\bibfnamefont {S.}~\bibnamefont {Gilardoni}}, \bibinfo
  {author} {\bibfnamefont {A.}~\bibnamefont {Drees}}, \bibinfo {author}
  {\bibfnamefont {W.}~\bibnamefont {Fischer}}, \bibinfo {author} {\bibfnamefont
  {S.}~\bibnamefont {Tepikian}}, \ and\ \bibinfo {author} {\bibfnamefont
  {S.~R.}\ \bibnamefont {Klein}},\ }\bibfield  {title} {\enquote {\bibinfo
  {title} {Observations of beam losses due to bound-free pair production in a
  heavy-ion collider},}\ }\href {\doibase 10.1103/PhysRevLett.99.144801}
  {\bibfield  {journal} {\bibinfo  {journal} {Physical Review Letters}\
  }\textbf {\bibinfo {volume} {99}},\ \bibinfo {pages} {144801} (\bibinfo
  {year} {2007})}\BibitemShut {NoStop}%
\bibitem [{\citenamefont {Weizs\"acker}(1934)}]{weizsacker_ausstrahlung_1934}%
  \BibitemOpen
  \bibfield  {author} {\bibinfo {author} {\bibfnamefont {C.~F.~v.}\
  \bibnamefont {Weizs\"acker}},\ }\bibfield  {title} {\enquote {\bibinfo
  {title} {Ausstrahlung bei {St\"oßen} sehr schneller {Elektronen}},}\ }\href
  {\doibase 10.1007/BF01333110} {\bibfield  {journal} {\bibinfo  {journal}
  {Zeitschrift für Physik}\ }\textbf {\bibinfo {volume} {88}},\ \bibinfo
  {pages} {612--625} (\bibinfo {year} {1934})}\BibitemShut {NoStop}%
\bibitem [{\citenamefont {Williams}(1934)}]{williams_nature_1934}%
  \BibitemOpen
  \bibfield  {author} {\bibinfo {author} {\bibfnamefont {E.~J.}\ \bibnamefont
  {Williams}},\ }\bibfield  {title} {\enquote {\bibinfo {title} {Nature of the
  {High} {Energy} {Particles} of {Penetrating} {Radiation} and {Status} of
  {Ionization} and {Radiation} {Formulae}},}\ }\href {\doibase
  10.1103/PhysRev.45.729} {\bibfield  {journal} {\bibinfo  {journal} {Physical
  Review}\ }\textbf {\bibinfo {volume} {45}},\ \bibinfo {pages} {729--730}
  (\bibinfo {year} {1934})}\BibitemShut {NoStop}%
\bibitem [{\citenamefont {Aste}\ \emph {et~al.}(1994)\citenamefont {Aste},
  \citenamefont {Hencken}, \citenamefont {Trautmann},\ and\ \citenamefont
  {Baur}}]{aste_electromagnetic_1994}%
  \BibitemOpen
  \bibfield  {author} {\bibinfo {author} {\bibfnamefont {Andreas}\ \bibnamefont
  {Aste}}, \bibinfo {author} {\bibfnamefont {Kai}\ \bibnamefont {Hencken}},
  \bibinfo {author} {\bibfnamefont {Dirk}\ \bibnamefont {Trautmann}}, \ and\
  \bibinfo {author} {\bibfnamefont {Gerhard}\ \bibnamefont {Baur}},\ }\bibfield
   {title} {\enquote {\bibinfo {title} {Electromagnetic pair production with
  capture},}\ }\href@noop {} {\bibfield  {journal} {\bibinfo  {journal}
  {Physical Review A}\ }\textbf {\bibinfo {volume} {50}},\ \bibinfo {pages}
  {3980--3983} (\bibinfo {year} {1994})}\BibitemShut {NoStop}%
\bibitem [{\citenamefont {Agger}\ and\ \citenamefont
  {S{\o}rensen}(1997)}]{agger_pair_1997}%
  \BibitemOpen
  \bibfield  {author} {\bibinfo {author} {\bibfnamefont {Carsten~K.}\
  \bibnamefont {Agger}}\ and\ \bibinfo {author} {\bibfnamefont {Allan~H.}\
  \bibnamefont {S{\o}rensen}},\ }\bibfield  {title} {\enquote {\bibinfo {title}
  {Pair creation with bound electron for photon impact on bare heavy nuclei},}\
  }\href {\doibase 10.1103/PhysRevA.55.402} {\bibfield  {journal} {\bibinfo
  {journal} {Physical Review A}\ }\textbf {\bibinfo {volume} {55}},\ \bibinfo
  {pages} {402--413} (\bibinfo {year} {1997})}\BibitemShut {NoStop}%
\bibitem [{\citenamefont {Belkacem}\ and\ \citenamefont
  {S{\o}rensen}(1998)}]{belkacem_bound-free_1998}%
  \BibitemOpen
  \bibfield  {author} {\bibinfo {author} {\bibfnamefont {Ali}\ \bibnamefont
  {Belkacem}}\ and\ \bibinfo {author} {\bibfnamefont {Allan~H.}\ \bibnamefont
  {S{\o}rensen}},\ }\bibfield  {title} {\enquote {\bibinfo {title} {Bound-free
  heavy-lepton pair production for photon and ion impact on atomic nuclei},}\
  }\href {\doibase 10.1103/PhysRevA.57.3646} {\bibfield  {journal} {\bibinfo
  {journal} {Physical Review A}\ }\textbf {\bibinfo {volume} {57}},\ \bibinfo
  {pages} {3646--3651} (\bibinfo {year} {1998})}\BibitemShut {NoStop}%
\bibitem [{\citenamefont {Aste}(2008)}]{aste_bound-free_2008}%
  \BibitemOpen
  \bibfield  {author} {\bibinfo {author} {\bibfnamefont {A.}~\bibnamefont
  {Aste}},\ }\bibfield  {title} {\enquote {\bibinfo {title} {Bound-free pair
  production cross-section in heavy-ion colliders from the equivalent photon
  approach},}\ }\href {\doibase 10.1209/0295-5075/81/61001} {\bibfield
  {journal} {\bibinfo  {journal} {EPL (Europhysics Letters)}\ }\textbf
  {\bibinfo {volume} {81}},\ \bibinfo {pages} {61001} (\bibinfo {year}
  {2008})}\BibitemShut {NoStop}%
\bibitem [{\citenamefont {Deneke}\ and\ \citenamefont
  {M\"uller}(2008)}]{deneke_bound-free_2008}%
  \BibitemOpen
  \bibfield  {author} {\bibinfo {author} {\bibfnamefont {C.}~\bibnamefont
  {Deneke}}\ and\ \bibinfo {author} {\bibfnamefont {C.}~\bibnamefont
  {M\"uller}},\ }\bibfield  {title} {\enquote {\bibinfo {title} {Bound-free
  $e^+e^-$ pair creation with a linearly polarized laser field and a nuclear
  field},}\ }\href {\doibase 10.1103/PhysRevA.78.033431} {\bibfield  {journal}
  {\bibinfo  {journal} {Physical Review A}\ }\textbf {\bibinfo {volume} {78}},\
  \bibinfo {pages} {033431} (\bibinfo {year} {2008})}\BibitemShut {NoStop}%
\bibitem [{\citenamefont {Artemyev}\ \emph {et~al.}(2012)\citenamefont
  {Artemyev}, \citenamefont {Jentschura}, \citenamefont {Serbo},\ and\
  \citenamefont {Surzhykov}}]{artemyev_boundfree_2012}%
  \BibitemOpen
  \bibfield  {author} {\bibinfo {author} {\bibfnamefont {A.~N.}\ \bibnamefont
  {Artemyev}}, \bibinfo {author} {\bibfnamefont {U.~D.}\ \bibnamefont
  {Jentschura}}, \bibinfo {author} {\bibfnamefont {V.~G.}\ \bibnamefont
  {Serbo}}, \ and\ \bibinfo {author} {\bibfnamefont {A.}~\bibnamefont
  {Surzhykov}},\ }\bibfield  {title} {\enquote {\bibinfo {title} {Bound–free
  pair production in ultra–relativistic ion collisions at the {LHC} collider:
  analytic approach to the total and differential cross sections},}\ }\href
  {\doibase 10.1140/epjc/s10052-012-1935-z} {\bibfield  {journal} {\bibinfo
  {journal} {The European Physical Journal C}\ }\textbf {\bibinfo {volume}
  {72}},\ \bibinfo {pages} {1935} (\bibinfo {year} {2012})}\BibitemShut
  {NoStop}%
\bibitem [{\citenamefont {Eichler}(1995)}]{eichler_relativistic_1995}%
  \BibitemOpen
  \bibfield  {author} {\bibinfo {author} {\bibfnamefont {J.}~\bibnamefont
  {Eichler}},\ }\href {\doibase 10.1016/B978-0-12-233675-1.X5023-4} {\emph
  {\bibinfo {title} {Relativistic {Atomic} {Collisions}}}}\ (\bibinfo
  {publisher} {Elsevier},\ \bibinfo {year} {1995})\BibitemShut {NoStop}%
\bibitem [{\citenamefont {Eichler}(2005)}]{eichler_lectures_2005}%
  \BibitemOpen
  \bibfield  {author} {\bibinfo {author} {\bibfnamefont {J}~\bibnamefont
  {Eichler}},\ }\bibfield  {title} {\enquote {\bibinfo {title} {Lectures on
  {Ion}-{Atom} {Collisions}},}\ }\href@noop {} {\bibfield  {journal} {\bibinfo
  {journal} {Lectures on Ion-Atom Collisions}\ } (\bibinfo {year}
  {2005})}\BibitemShut {NoStop}%
\bibitem [{\citenamefont {Rose}(1957)}]{rose_elementary_1957}%
  \BibitemOpen
  \bibfield  {author} {\bibinfo {author} {\bibfnamefont {Morris~Edgar}\
  \bibnamefont {Rose}},\ }\href@noop {} {\emph {\bibinfo {title} {Elementary
  theory of angular momentum}}}\ (\bibinfo  {publisher} {Wiley},\ \bibinfo
  {year} {1957})\BibitemShut {NoStop}%
\bibitem [{\citenamefont {Pratt}(1960)}]{pratt_atomic_1960}%
  \BibitemOpen
  \bibfield  {author} {\bibinfo {author} {\bibfnamefont {R.~H.}\ \bibnamefont
  {Pratt}},\ }\bibfield  {title} {\enquote {\bibinfo {title} {Atomic
  {Photoelectric} {Effect} at {High} {Energies}},}\ }\href {\doibase
  10.1103/PhysRev.117.1017} {\bibfield  {journal} {\bibinfo  {journal}
  {Physical Review}\ }\textbf {\bibinfo {volume} {117}},\ \bibinfo {pages}
  {1017--1028} (\bibinfo {year} {1960})}\BibitemShut {NoStop}%
\bibitem [{\citenamefont {Alling}\ and\ \citenamefont
  {Johnson}(1965)}]{alling_exact_1965}%
  \BibitemOpen
  \bibfield  {author} {\bibinfo {author} {\bibfnamefont {W.~R.}\ \bibnamefont
  {Alling}}\ and\ \bibinfo {author} {\bibfnamefont {W.~R.}\ \bibnamefont
  {Johnson}},\ }\bibfield  {title} {\enquote {\bibinfo {title} {Exact
  {Calculation} of {K}-{Shell} and {L}-{Shell} {Photoeffect}},}\ }\href
  {\doibase 10.1103/PhysRev.139.A1050} {\bibfield  {journal} {\bibinfo
  {journal} {Physical Review}\ }\textbf {\bibinfo {volume} {139}},\ \bibinfo
  {pages} {A1050--A1062} (\bibinfo {year} {1965})}\BibitemShut {NoStop}%
\bibitem [{\citenamefont {Ichihara}\ \emph {et~al.}(1996)\citenamefont
  {Ichihara}, \citenamefont {Shirai},\ and\ \citenamefont
  {Eichler}}]{ichihara_radiative_1996}%
  \BibitemOpen
  \bibfield  {author} {\bibinfo {author} {\bibfnamefont {A.}~\bibnamefont
  {Ichihara}}, \bibinfo {author} {\bibfnamefont {T.}~\bibnamefont {Shirai}}, \
  and\ \bibinfo {author} {\bibfnamefont {J\"org}\ \bibnamefont {Eichler}},\
  }\bibfield  {title} {\enquote {\bibinfo {title} {Radiative electron capture
  and the photoelectric effect at high energies},}\ }\href {\doibase
  10.1103/PhysRevA.54.4954} {\bibfield  {journal} {\bibinfo  {journal}
  {Physical Review A}\ }\textbf {\bibinfo {volume} {54}},\ \bibinfo {pages}
  {4954--4959} (\bibinfo {year} {1996})}\BibitemShut {NoStop}%
\bibitem [{\citenamefont {Sodickson}\ \emph {et~al.}(1961)\citenamefont
  {Sodickson}, \citenamefont {Bowman}, \citenamefont {Stephenson},\ and\
  \citenamefont {Weinstein}}]{sodickson_single-quantum_1961}%
  \BibitemOpen
  \bibfield  {author} {\bibinfo {author} {\bibfnamefont {L.}~\bibnamefont
  {Sodickson}}, \bibinfo {author} {\bibfnamefont {W.}~\bibnamefont {Bowman}},
  \bibinfo {author} {\bibfnamefont {J.}~\bibnamefont {Stephenson}}, \ and\
  \bibinfo {author} {\bibfnamefont {R.}~\bibnamefont {Weinstein}},\ }\bibfield
  {title} {\enquote {\bibinfo {title} {Single-{Quantum} {Annihilation} of
  {Positrons}},}\ }\href {\doibase 10.1103/PhysRev.124.1851} {\bibfield
  {journal} {\bibinfo  {journal} {Physical Review}\ }\textbf {\bibinfo {volume}
  {124}},\ \bibinfo {pages} {1851--1861} (\bibinfo {year} {1961})}\BibitemShut
  {NoStop}%
\bibitem [{\citenamefont {Grant}(2007)}]{grant_relativistic_2007}%
  \BibitemOpen
  \bibfield  {author} {\bibinfo {author} {\bibfnamefont {Ian~P.}\ \bibnamefont
  {Grant}},\ }\href@noop {} {\emph {\bibinfo {title} {Relativistic {Quantum}
  {Theory} of {Atoms} and {Molecules}: {Theory} and {Computation}}}},\ Springer
  {Series} on {Atomic}, {Optical}, and {Plasma} {Physics}\ (\bibinfo
  {publisher} {Springer-Verlag},\ \bibinfo {address} {New York},\ \bibinfo
  {year} {2007})\BibitemShut {NoStop}%
\bibitem [{\citenamefont {Johansson}(2017)}]{johansson_arb:_2017}%
  \BibitemOpen
  \bibfield  {author} {\bibinfo {author} {\bibfnamefont {F.}~\bibnamefont
  {Johansson}},\ }\bibfield  {title} {\enquote {\bibinfo {title} {Arb:
  efficient arbitrary-precision midpoint-radius interval arithmetic},}\ }\href
  {\doibase 10.1109/TC.2017.2690633} {\bibfield  {journal} {\bibinfo  {journal}
  {IEEE Transactions on Computers}\ }\textbf {\bibinfo {volume} {66}},\
  \bibinfo {pages} {1281--1292} (\bibinfo {year} {2017})}\BibitemShut {NoStop}%
\bibitem [{\citenamefont {{Bhabha H. J.}}\ \emph {et~al.}(1934)\citenamefont
  {{Bhabha H. J.}}, \citenamefont {{Hulme H. R.}},\ and\ \citenamefont {{Fowler
  Ralph Howard}}}]{bhabha_h._j._annihilation_1934}%
  \BibitemOpen
  \bibfield  {author} {\bibinfo {author} {\bibnamefont {{Bhabha H. J.}}},
  \bibinfo {author} {\bibnamefont {{Hulme H. R.}}}, \ and\ \bibinfo {author}
  {\bibnamefont {{Fowler Ralph Howard}}},\ }\bibfield  {title} {\enquote
  {\bibinfo {title} {The annihilation of fast positrons by electrons in the
  {K}-shell},}\ }\href {\doibase 10.1098/rspa.1934.0184} {\bibfield  {journal}
  {\bibinfo  {journal} {Proc. R. Soc. London, Ser. A}\ }\textbf
  {\bibinfo {volume} {146}},\ \bibinfo {pages} {723--736} (\bibinfo {year}
  {1934})}\BibitemShut {NoStop}%
\bibitem [{\citenamefont {Johnson}(1967)}]{johnson_angular_1967}%
  \BibitemOpen
  \bibfield  {author} {\bibinfo {author} {\bibfnamefont {W.~R.}\ \bibnamefont
  {Johnson}},\ }\bibfield  {title} {\enquote {\bibinfo {title} {Angular
  {Distribution} of {Single}-{Quantum} {Annihilation} {Radiation}},}\ }\href
  {\doibase 10.1103/PhysRev.159.61} {\bibfield  {journal} {\bibinfo  {journal}
  {Physical Review}\ }\textbf {\bibinfo {volume} {159}},\ \bibinfo {pages}
  {61--68} (\bibinfo {year} {1967})}\BibitemShut {NoStop}%
\end{thebibliography}%
	
\end{document}